# FORMATION OF MICROCRYSTALLINE STRUCTURE IN LARGE-SCALE INGOTS OF Ti ALLOY TI-6AL-4V DURING THE COMPLEX LOADING

V. K. Berdin, M. V. Karavaeva, S. K. Kiseleva

*Ufa state aviation technical university, K.Marks Str.12, 450000, Ufa, Russia*

A new method of combined loading was developed for manufacturing large - scale ingot of alpha-beta titanium alloy Ti-6Al-4V with the diameter of the gauge 120 mm and the length 300 mm. The process includes torsion with simultaneous tensile deformation, followed by a single compression and drawing of the ingot in alpha-beta temperature phase fields. Ingot structure and mechanical property evaluation was performed after deformation and post-deformation heating. The ingots had completely homogeneous macrostructure. The ingot microstructures were alpha plus beta with a mean grain size of $d_\alpha=10$ μm. There has not been any significant difference in microstructure of central and periphery areas in the cross section of the ingot.

**Introduction**

The efficiency of the transformation process of an initial coarse-grained structure into the microcrystalline one in two-phase Ti alloys depends on various parameters of thermomechanical material treatment. The main ones are temperature, strain rate and degree of the accumulated strain [1].

One of the most important factors affecting the formation of the microcrystalline structure is a loading schedule: the combination and the sequencing of deformation methods. The loading schedule provides the achievement of the deformation degree e in the whole deformed volume, which is necessary for developing and completing the recrystallization processes. According to the authors of the papers [1-5], this level can be determined by the material properties, the initial microstructure and, as a rule, exceeds the value $e = 0.7-1.1$.

In papers [2-6] the influence of the schedule of one- and two-component loading on the development of lamellar microstructure transformation in two-phase titanium alloys into the equiaxial one has been studied. It has been pointed out that from the point of view of obtaining the homogeneous deformation state in the bulk of the material, the schedules of uniaxial tension or compression at the absolute absence of any friction between the ingot and the tool, are suggested to be the best ones. Nevertheless, under tension of the cylindrical sample, the plastic flow of the material is being localized at the deformation degree $e = 0.2-0.4$, which leads to necking in the gauge of the sample, where the microstructure alterations start focusing [4].

During the deformation of the cylindrical sample by torsion, the value of the accumulated deformation in the gauge of the sample is inhomogeneous. In the direction of the axis of symmetry of the sample, the deformation degree stays constant. At the same time, along the sample radius the deformation degree increases linear from zero in the centre to its maximum at the lateral surface in accordance with the flat cross-section hypothesis [7]. The microstructure changes are being developed in the sample in the same way: they are insignificant in the centre, while there is a considerable refinement of the initial coarse-grained structure on the periphery [5].

During the two-component (torsion + tension) proportional loading the gradient of the deformation degree in the volume of the sample does not disappear, but it is still less, than the one during the pure torsion. At the same time, even the small share of the axial deformation added to the torsion component increases the homogeneity of the transformation process development of the microstructure in the cross section of the sample [4, 6]. The following uniaxial tension absolutely equalizes the average grain size and the phase composition in the whole deformed volume [2-4].

The carried out experiments at the laboratory level [2-5] have shown the possibility in principal to use two-component loading as the method, forming the specified microcrystalline structure in cylindrical samples from the two-phase Ti alloys, which have a coarse-grained microstructure with lamellar phase morphology. The following thermomechanical deformation modes have been determined: the loading schedule – torsion, combined with tension (or compression), with the following uniaxial tension (or compression) and drawing in the range of temperatures $T=T_{ps}$ -50 °C and the strain rates $\xi=10^{-3}-10^{-2}$ s$^{-1}$ up to the total strain degree e >2.0.

This paper presents the example of the further development of the method of a coarse-grained microstructure refinement under conditions of complex loading, based on the hot-deformation process of long-length cylindrical ingots by torsion with simultaneous tension, as applied to large scale ingots [8].

**Experimental procedure**

The ingot of Ti-6Al-4V alloy has been used as the basic material here. To equalize the chemical composition of the ingot before the beginning of the strain they have been carrying out a homogenizing annealing at 1100 °C for 5 hours. For a preliminary refinement of β - grains they have carried out forging of ingots on the hydraulic press in isothermal die set at temperature of one-phase β-area in the following mode: the ingot heating up to 1100 °C, temperature of the block heads 900 °C, drawing for a billet with a square cross section with an edge 200 mm, air cooling; the ingot reheating up to 1100 °C, temperature of the block heads 900 °C, drawing for a billet with a diameter of a cross section 120 mm with the following air cooling.

On the base of the analysis of the previous results [2-5,8,9] the following temperature-strain rate modes of deformation ingot treatment from Ti-6Al-4V alloy have been chosen: the temperature of treatment 920 °C ($T_{ps}$ -50 °C) and the speed of torsion of the mobile traverse of the testing machine ω=1.0 rotations/min.

The full experiment has been carried out on ingots with the diameter of the gauge 120 mm and the length 300 mm. The loading schedule included torsion with simultaneous tensile deformation, followed by a single compression and drawing of the ingot to the initial size.

The torsion with simultaneous tension has been carried out on the drawing mill. The gauge of the ingot had been under isothermal conditions. The ingot has been deformed on the degree of final strain e on the outer surface equal to 4.3 (N=6 turns, ΔL=35 mm).

Then, a deformed piece had been cut off the ingot, heated in the furnace up to 920 °C and deformed by compression up to the deformation strain by height 40-50% followed by the drawing to the initial size along the ingot axis on the flat heads in isothermal die set, using the hydraulic press.

Annealing has been made after deformation according to the mode: $T_1$=850 °C with holding during 60 min., cooling with furnace up to $T_2$=750 °C with holding during 30 min, and air cooling up to room temperature.

Microstructure was studied using a «JSM-6390» scanning electron microscope (SEM) and «JEM-2100» transmission electron microscope (TEM). The microstructure was examined in a section, perpendicular to the axis of specimen near the outer surface (zone of maximum deformation) and in the central part. The following parameters were determined: the mean grain size $d_\alpha$ of alpha-phase particles; the mean value of the particle shape coefficient $K_\alpha$ ($K_\alpha$ – a ratio length of particle alpha-phase to its thickness) and a volume fraction of alpha-phase $V_\alpha$ (%).Quantitative analysis of microstructure was carried out using an optical microscope «Axiotech 100» with adapter of quantitative image analysis KS300.

The analysis of a crystallographic texture has been carried out on the samples cut out in the direction, which coincides with the axis of symmetry of the initial ingot. The crystallographic texture has been studied by the X-ray diffraction method on the diffractometer «Dron-3M» using the tube with copper radiation.

The mechanical properties of the material have been determined in the result of static and dynamical tests at room temperature on standard samples, cut out in the axial direction. The samples for uniaxial tension had the gauge diameter 5,0 mm and the length 25,0 mm. The samples for impact bending tests in the design section had the size 10,0 x 8,0mm.

**Results**

The microstructure of the as-received materials consists of primary beta-grains with mean grain size 1.5-2.0 mm. Inside the primary β-grains in the cooling process, the microstructure of a lamellar kind has been formed from a one-phase area, with the plate thickness of α-phase equal to 2-3 μm (Fig. 1, a) and with the evident α-phase along the boundaries of the primary β-grains. Electron microscopic study of thin foils has pointed out that there are both separate dislocations and a few dislocation walls inside the α-plates (Fig. 1, b).

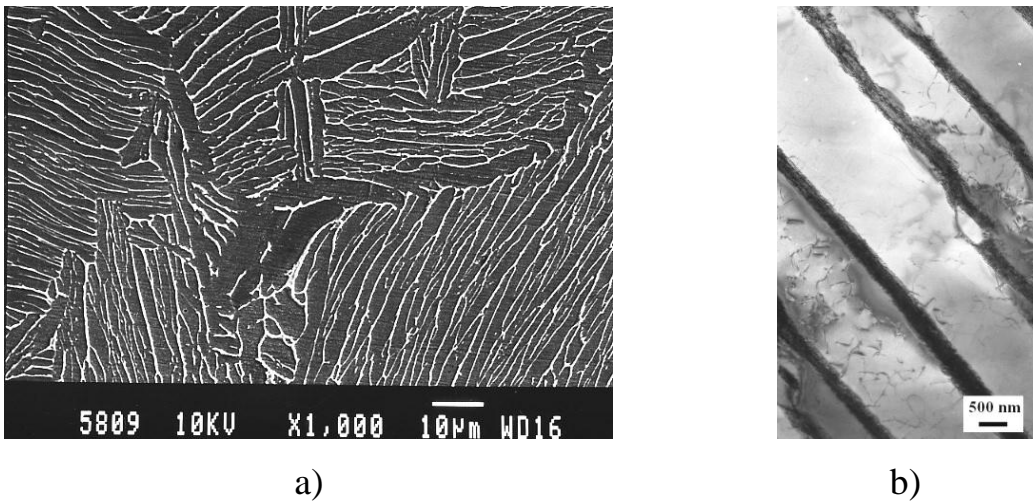

a) b)

Fig. 1. Microstructure of the as-received materials:
a – scanning electron microscopy; b – transmission electron microscopy

The appearance of an ingot after deformation processing has been presented in Fig. 2. The traces of deformation macrolocalization at the sample gauge have not been revealed. The angle of the fibers' twist on the ingot surface is constant along the whole length of the gauge, which testifies the homogeneous development of plastic deformation of material along the ingot axis.

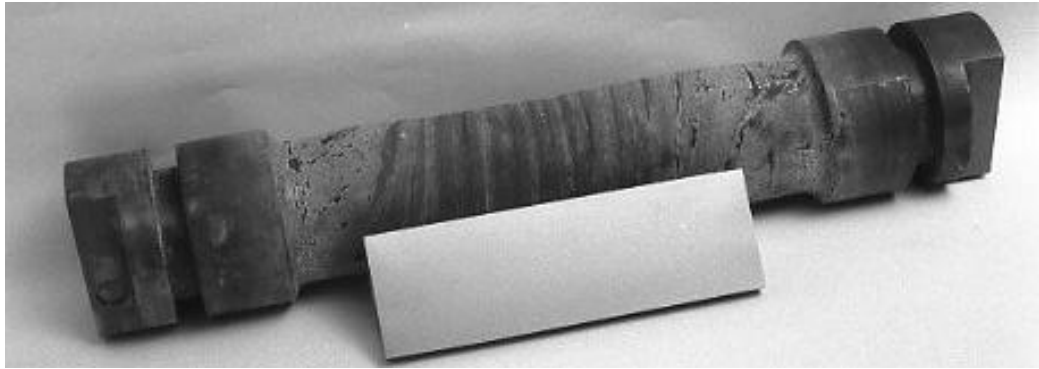

Fig 2. The appearance of a large scale ingot out of the alloy Ti-6Al-4V after torsion with simultaneous tension and macrostructure of the ingot gauge after the complex loading

The macrostructure of the ingots obtained by deformation according to the schedule of complex loading in the longitudinal section is homogeneous, without recrystallized β-grains. The initial coarse-grained microstructure with lamellar phase morphology in the deformation process has absolutely transformed into microcrystalline equiaxial one with the coefficient of the grain shape of the α-phase K=2.0. The grain size of the α-phase calculated according to the phase boundaries has comprised d= 10 μm. The microstructure is presented in Fig. 3, a. There has not been any significant difference in microstructure of central and periphery areas in the cross section of the ingot. The carried out research of the fine structure by a transmission electron microscope has pointed out that the grains of α-phase, observed at less magnification in a scanning electron microscope are divided by the grain boundaries and characterized by a high dislocation density (Fig. 3, b).

Thermal treatment provided termination of the recrystallization processes, which was followed by development of the duplex type structure. Intercrystalline and phase boundaries righted and α-phase grains became almost free from dislocations (Fig. 3, b, c). The average grain size, after the thermal treatment has been calculated, made up d=2.0 μm according to intercrystalline boundaries in the pictures, which have been made on the transmission electron microscope. At the same time there hasn't been any significant difference between the grain size in longitudinal and cross sections.

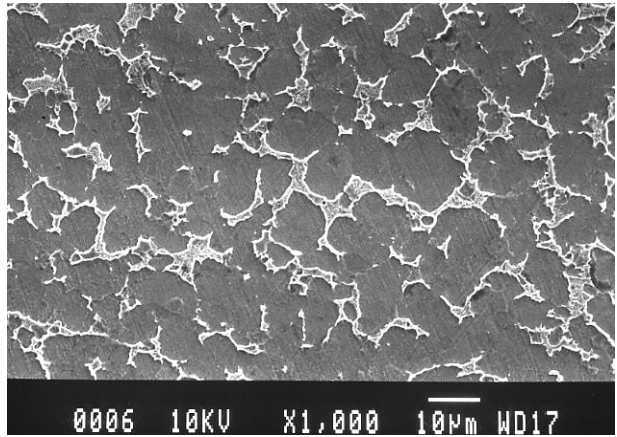 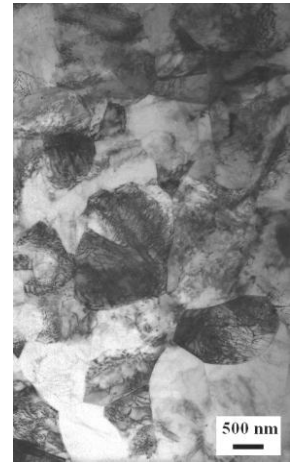

a)                          b)

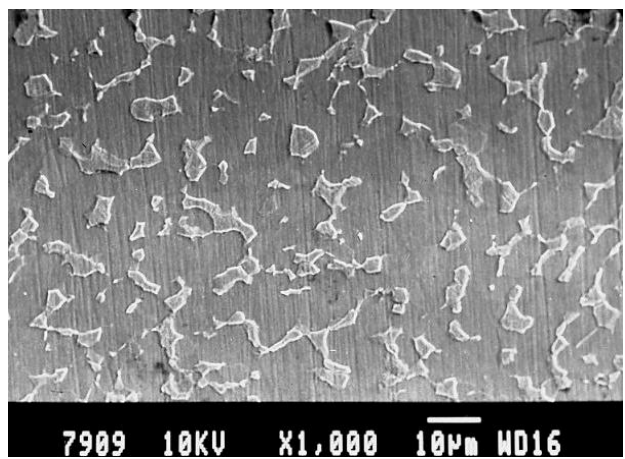 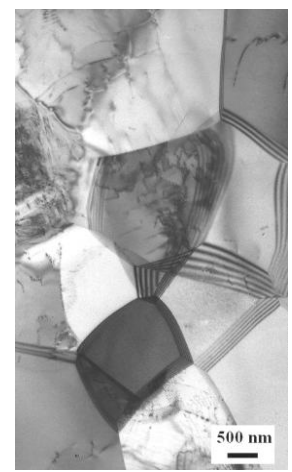

c)                          d)

Fig. 3. Microstructure of the gauge of a large scale ingot of the alloy Ti-6Al-4V: a, b – after deformation processing; c, d – after annealing; a, c – scanning electron microscopy; b, d- transmission electron microscopy

The results of the X-ray diffraction method are presented in Fig. 4 as the reverse pole figure, where the intensity of the pole density of normal lines for 11 planes has been shown. The reverse pole figure shows that there is no any preferable orientation of any definite plane in relation to the axis of the ingot. So, the thermomechanical treatment according to the scheme of complex loading leads to obtaining an almost non-textural state in the material.

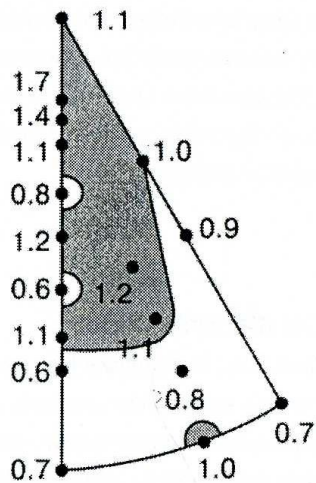

Fig. 4. The reverse pole figure for cross section of large scale ingot of the alloy Ti-6Al-4V after deformation according to the complex loading schedule

The level of the obtained mechanical properties (Table 1) in the ingot material with the obtained microcrystalline structure fully satisfies the claims for roll forgings with the diameter >50.0 mm, and does not comply to mechanical properties of this alloy with microcrystalline structure with the average grain size d=10.0 μm, processed by the multiaxial forging (according to the scanning electron microscope data) [10].

Table 1. Mechanical properties of the material of coarse-grained ingot out of the alloy Ti-6Al-4V after hot deformation due to the complex loading schedule

| $\sigma_u$, MPa | $\sigma_y$, MPa | $\delta$, % | $\psi$, % | KCU, J/cm$^2$ | HRC | The information source |
|---|---|---|---|---|---|---|
| 1024.0 | 965.1 | 17.3 | 44.0 | 43.2 | 35 | This research* |
| 962.0 | 910.0 | 15.4 | 29.8 | - | - | [10] |

*The average values are presented here according to the testing results of four samples.

**Concluding remarks**

The carried out research of the combined loading as the microstructure refinement method in a large scale ingot of Ti alloy Ti-6Al-4V has shown a crucial possibility to use torsion with extra tension at least as an initial stage of preparation of the microcrystalline structure in two-phase titanium alloys.

The application of the torsion at the first stage of the hot deformation treatment allows plastify the outer ingot layers due to transformation of coarse-grained structure into microcrystalline one. After this the additional operations of compression or tension may be used to work out the central parts of axisymmetric ingot. Such an approach can provide a considerable reduction of costs of obtaining large capacity axisymmetryic ingots out of two-phase titanium alloys with the specified distributed in the bulk of a microcrystalline structure.